\documentclass{Rinton-P10x7}

\def\beq{\begin{equation}}
\def\eeq{\end{equation}}
\def\beqa{\begin{eqnarray}}
\def\eeqa{\end{eqnarray}}

\def\za{\alpha}
\def\zb{\beta}
\def\lsim{\mathrel{\raise.3ex\hbox{$<$\kern-.75em\lower1ex\hbox{$\sim$}}} }
\def\gsim{\mathrel{\raise.3ex\hbox{$>$\kern-.75em\lower1ex\hbox{$\sim$}}} }

\begin{document}
\thispagestyle{empty}

\onecolumn

\begin{flushright}
IPAS-HEP-k014\\
Apr 2001
\end{flushright}

\vspace*{.5in}

\begin{center}
{\bf  \Large Some Recent Results from the Generic Supersymmetric
Standard Model $^\star$ }\\
\vspace*{.5in}
{\bf  Otto C.W. Kong}\\[.05in]
{\it Institute of Physics, Academia Sinica, Nankang, Taipei, TAIWAN 11529}

\vspace*{.8in}
{Abstract}\\
\end{center}

The generic supersymmetric standard model is a model built from a 
supersymmetrized standard model field spectrum the gauge symmetries only. The 
popular minimal supersymmetric standard model differs from the generic version in 
having R-parity imposed by hand. We review an efficient formulation of the model
and some of the recently obtained interesting phenomenological features.
The latter includes R-parity violating contributions to scalar masses that
had been largely overlooked and the related contributions to fermion
electric dipole moments and $\mu \to e\,\gamma$.

\vfill
\noindent --------------- \\
$^\star$ Talk  presented  at PASCOS 2001 (Apr 10-15), Chapel Hill, NC USA\\
 --- submission for the proceedings.  
 
\clearpage
\addtocounter{page}{-1}

\title{Some recent results from the Generic Supersymmetric
Standard Model}

\author{Otto C. W. Kong}

\address{Institute of Physics, Academia Sinica, Nankang, Taipei, TAIWAN 11529\\E-mail: kongcw@phys.sinica.edu.tw}

\maketitle

\abstracts{
The generic supersymmetric standard model is a model built from a 
supersymmetrized standard model field spectrum the gauge symmetries only. The 
popular minimal supersymmetric standard model differs from the generic version in 
having R-parity imposed by hand. We review an efficient formulation of the model
and some of the recently obtained interesting phenomenological features.
The latter includes R-parity violating contributions to scalar masses that
had been largely overlooked and the related contributions to fermion
electric dipole moments and $\mu \to e\,\gamma$.
}

\section{The Generic Supersymmetric Standard Model}
A theory built with the minimal superfield spectrum incorporating the Standard 
Model (SM) particles and interactions dictated by the SM (gauge) symmetries, and 
the idea that supersymmetry (SUSY) is softly broken is what should be called the
the generic supersymmetric standard model. The popular minimal 
supersymmetric standard model differs from the generic version in having
a discrete symmetry, called R parity, imposed by hand to enforce baryon and
lepton number conservation. With the strong experimental hints at the existence
of lepton number violating neutrino masses, such a theory of SUSY without R-parity
deserves more attention than ever before. The generic supersymmetric standard model
contains all kind of (so-called) R-parity violating (RPV) parameters.
The latter includes the more popular trilinear ($\lambda_{ijk}$, $\lambda_{ijk}^{\prime}$, and	$\lambda_{ijk}^{\prime\prime}$) and bilinear ($\mu_i$)
couplings in the superpotential, as well as  soft SUSY breaking
parameters of the trilinear, bilinear, and soft mass (mixing) types. In order not 
to miss any plausible RPV phenomenological features, it is important that all of 
the RPV parameters be taken into consideration without {\it a priori} bias. 
We do, however, expect some sort of symmetry principle to guard against
the dangerous proton decay problem. The emphasis is hence put on the lepton number
violating phenomenology.

The renormalizable superpotential for the generic supersymmetric standard model
can be written  as
\small\beqa
W \!\! &=& \!\varepsilon_{ab}\Big[ \mu_{\alpha}  \hat{H}_u^a \hat{L}_{\alpha}^b 
+ h_{ik}^u \hat{Q}_i^a   \hat{H}_{u}^b \hat{U}_k^{\scriptscriptstyle C}
+ \lambda_{\alpha jk}^{\!\prime}  \hat{L}_{\alpha}^a \hat{Q}_j^b
\hat{D}_k^{\scriptscriptstyle C} 
+
\frac{1}{2}\, \lambda_{\alpha \beta k}  \hat{L}_{\alpha}^a  
 \hat{L}_{\beta}^b \hat{E}_k^{\scriptscriptstyle C} \Big] + 
\frac{1}{2}\, \lambda_{ijk}^{\!\prime\prime}  
\hat{U}_i^{\scriptscriptstyle C} \hat{D}_j^{\scriptscriptstyle C}  
\hat{D}_k^{\scriptscriptstyle C}   ,
\eeqa\normalsize
where  $(a,b)$ are $SU(2)$ indices, $(i,j,k)$ are the usual family (flavor) 
indices, and $(\za, \zb)$ are extended flavor indices going from $0$ to $3$.
At the limit where $\lambda_{ijk}, \lambda^{\!\prime}_{ijk},  
\lambda^{\!\prime\prime}_{ijk}$ and $\mu_{i}$  all vanish, 
one recovers the expression for the R-parity preserving case, 
with $\hat{L}_{0}$ identified as $\hat{H}_d$. Without R-parity imposed,
the latter is not {\it a priori} distinguishable from the $\hat{L}_{i}$'s.
Note that $\lambda$ is antisymmetric in the first two indices, as
required by  the $SU(2)$  product rules, as shown explicitly here with 
$\varepsilon_{\scriptscriptstyle 12} =-\varepsilon_{\scriptscriptstyle 21}=1$.
Similarly, $\lambda^{\!\prime\prime}$ is antisymmetric in the last two 
indices, from $SU(3)_{\scriptscriptstyle C}$. 

The soft SUSY breaking part 
of the Lagrangian is more interesting, if only for the fact that  many
of its interesting details have been overlooked in the literature.
However, we will postpone the discussion till after we address the
parametrization issue.\\[-.2in]

\section{Parametrization}
Doing phenomenological studies without specifying a choice 
of flavor bases is ambiguous. It is like doing SM quark physics with 18
complex Yukawa couplings, instead of the 10 real physical parameters.
As far as the SM itself is concerned, the extra 26 real parameters
are simply redundant, and attempts to relate the full 36 parameters to
experimental data will be futile.
In the generic supersymmetric standard model, the choice of an optimal
parametrization mainly concerns the 4 $\hat{L}_\alpha$ flavors. We use
here the single-VEV parametrization\cite{ru,as8} (SVP), in which flavor bases 
are chosen such that : 
1/ among the $\hat{L}_\alpha$'s, only  $\hat{L}_0$, bears a VEV,
{\it i.e.} {\small $\langle \hat{L}_i \rangle \equiv 0$};
2/  {\small $h^{e}_{jk} (\equiv \lambda_{0jk}) 
=\frac{\sqrt{2}}{v_{\scriptscriptstyle 0}} \,{\rm diag}
\{m_{\scriptscriptstyle 1},
m_{\scriptscriptstyle 2},m_{\scriptscriptstyle 3}\}$};
3/ {\small $h^{d}_{jk} (\equiv \lambda^{\!\prime}_{0jk} =-\lambda_{j0k}) 
= \frac{\sqrt{2}}{v_{\scriptscriptstyle 0}}{\rm diag}\{m_d,m_s,m_b\}$}; 
4/ {\small $h^{u}_{ik}=\frac{\sqrt{2}}{v_{\scriptscriptstyle u}}
V_{\mbox{\tiny CKM}}^{\!\scriptscriptstyle T} \,{\rm diag}\{m_u,m_c,m_t\}$}, where 
${v_{\scriptscriptstyle 0}} \equiv  \sqrt{2}\,\langle \hat{L}_0 \rangle$
and ${v_{\scriptscriptstyle u} } \equiv \sqrt{2}\,
\langle \hat{H}_{u} \rangle$. The big advantage of the SVP is that it gives 
the complete tree-level mass matrices of all the states (scalars and fermions) the simplest structure\cite{as5,as8}.

\section{Fermion Sector Phenomenology}
The SVP gives quark mass matrices exactly in the SM form. For the masses
of the color-singlet fermions, all the RPV effects are paramatrized by the
$\mu_i$'s only. For example, the five charged fermions ( gaugino
+ Higgsino + 3 charged leptons ), we have
\small\beq \label{mc}
{\mathcal{M}_{\scriptscriptstyle C}} =
 \left(
{\begin{array}{ccccc}
{M_{\scriptscriptstyle 2}} &  
\frac{g_{\scriptscriptstyle 2}{v}_{\scriptscriptstyle 0}}{\sqrt 2}  
& 0 & 0 & 0 \\
 \frac{g_{\scriptscriptstyle 2}{v}_{\scriptscriptstyle u}}{\sqrt 2} & 
 {{ \mu}_{\scriptscriptstyle 0}} & {{ \mu}_{\scriptscriptstyle 1}} &
{{ \mu}_{\scriptscriptstyle 2}}  & {{ \mu}_{\scriptscriptstyle 3}} \\
0 &  0 & {{m}_{\scriptscriptstyle 1}} & 0 & 0 \\
0 & 0 & 0 & {{m}_{\scriptscriptstyle 2}} & 0 \\
0 & 0 & 0 & 0 & {{m}_{\scriptscriptstyle 3}}
\end{array}}
\right)  \; .
\eeq\normalsize
Moreover each $\mu_i$ parameter here characterizes directly the RPV effect
on the corresponding charged lepton  ($\ell_i = e$, $\mu$, and $\tau$).
This, and the corresponding neutrino-neutralino masses and mixings,
has been exploited to implement a detailed study of the tree-level
RPV phenomenology from the gauge interactions, with interesting 
results\cite{ru}.

Neutrino masses and oscillations is no doubt one of the most important aspects
of the model. Here, it is particularly important that the various RPV contributions 
to neutrino masses, up to 1-loop level, be studied in a framework that takes no 
assumption on the other parameters. Our formulation provides such a framework.
Interested readers are referred to Refs.\cite{ok,as1,as5,as9,AL}.

\section{SUSY Breaking Terms and Related Phenomenology}

Obtaining the squark and slepton masses is straightforward, once all the 
admissible soft SUSY breaking terms are explicitly down\cite{as5}. The only
RPV contribution to the squark masses\cite{as5,as4} is given by a
$- (\, \mu_i^*\lambda^{\!\prime}_{ijk}\, ) \; 
\frac{v_{\scriptscriptstyle u}}{\sqrt{2}}$ term in the $LR$ mixing part.
Note that the term contains flavor-changing ($j\ne k$) parts which,
unlike the $A$-terms ones, cannot be suppressed through a flavor-blind
SUSY breaking spectrum. Hence, it has very interesting implications
to quark electric dipole moments (EDM's) and related processses
such as $b\to s\, \gamma$\cite{as4,as6,kk,cch1}. For instance, it contributes
to neutron EDM at 1-loop order, through a simple gluino diagram of the
$d$ squark. If one naively imposes the constraint for this RPV contribution 
itself not to exceed the experimental bound on neutron EDM, one gets roughly
$\mbox{Im}(\mu_i^*\lambda^{\!\prime}_{i\scriptscriptstyle 1\!1}) 
\lsim 10^{-6}\,\mbox{GeV}$, a constraint that is interesting even
in comparison to the bounds on the corresponding parameters obtainable
from asking no neutrino masses to exceed the super-Kamiokande 
atmospheric oscillation scale\cite{as4}. We have performed an extensive 
analytical and numerical study, including also the charginolike and neutralinolike 
contributions, to the neutron EDM\cite{as6}. The parameter combination
$\mu_i^*\lambda^{\!\prime}_{i\scriptscriptstyle 1\!1}$ is shown to be 
well constrained, with little dependence on $\tan\!\zb$. This applies not only
to the complex phase, or imaginary part of, the combination. Real 
$\mu_i^*\lambda^{\!\prime}_{i\scriptscriptstyle 1\!1}$, in the presence of
complex phases in the gaugino and Higgsino mass parameters, also contributes
to neutron EDM. We refer readers to the reference for discussions on various
novel features illustrated by the detailed study. 

The mass matrices are a bit more complicated in the scalar sectors\cite{as5,as7}.
The $1+4+3$ charged scalar masses are given in terms of the blocks
\small\beqa
&& \widetilde{\cal M}_{\!\scriptscriptstyle H\!u}^2 =
\tilde{m}_{\!\scriptscriptstyle H_{\!\scriptscriptstyle u}}^2
+ \mu_{\!\scriptscriptstyle \za}^* \mu_{\scriptscriptstyle \za}^{}
+ M_{\!\scriptscriptstyle Z}^2\, \cos\!2 \beta 
\left[ \,\frac{1}{2} - \sin\!^2\theta_{\!\scriptscriptstyle W}\right]
+ M_{\!\scriptscriptstyle Z}^2\,  \sin\!^2 \beta \;
[1 - \sin\!^2 \theta_{\!\scriptscriptstyle W}]
\; ,
\nonumber \\
&&\widetilde{\cal M}_{\!\scriptscriptstyle LL}^2
= \tilde{m}_{\!\scriptscriptstyle {L}}^2 +
m_{\!\scriptscriptstyle L}^\dag m_{\!\scriptscriptstyle L}^{}
+ M_{\!\scriptscriptstyle Z}^2\, \cos\!2 \beta 
\left[ -\frac{1}{2} +  \sin\!^2 \theta_{\!\scriptscriptstyle W}\right] 
+ \left( \begin{array}{cc}
 M_{\!\scriptscriptstyle Z}^2\,  \cos\!^2 \beta \;
[1 - \sin\!^2 \theta_{\!\scriptscriptstyle W}] 
& \quad 0_{\scriptscriptstyle 1 \times 3} \quad \\
0_{\scriptscriptstyle 3 \times 1} & 0_{\scriptscriptstyle 3 \times 3}  
\end{array} \right) 
+ (\mu_{\!\scriptscriptstyle \za}^* \mu_{\scriptscriptstyle \zb}^{})
\; ,
\nonumber \\
&& \widetilde{\cal M}_{\!\scriptscriptstyle RR}^2 =
\tilde{m}_{\!\scriptscriptstyle {E}}^2 +
m_{\!\scriptscriptstyle E}^{} m_{\!\scriptscriptstyle E}^\dag
+ M_{\!\scriptscriptstyle Z}^2\, \cos\!2 \beta 
\left[  - \sin\!^2 \theta_{\!\scriptscriptstyle W}\right] \; ; \qquad
\eeqa
{\normalsize and}
\beqa 
\label{ELH}
\widetilde{\cal M}_{\!\scriptscriptstyle LH}^2
&=& (B_{\za}^*)  
+ \left( \begin{array}{c} 
{1 \over 2} \,
M_{\!\scriptscriptstyle Z}^2\,  \sin\!2 \beta \,
[1 - \sin\!^2 \theta_{\!\scriptscriptstyle W}]  \\
0_{\scriptscriptstyle 3 \times 1} 
\end{array} \right)\; ,
\qquad
\\
\label{ERH}
\widetilde{\cal M}_{\!\scriptscriptstyle RH}^2
&=&  -\,(\, \mu_i^*\lambda_{i{\scriptscriptstyle 0}k}\, ) \; 
\frac{v_{\scriptscriptstyle 0}}{\sqrt{2}} \; ,
\\ 
\label{ERL}
(\widetilde{\cal M}_{\!\scriptscriptstyle RL}^{2})^{\scriptscriptstyle T} 
&=& \left(\begin{array}{c} 
0  \\   A^{\!{\scriptscriptstyle E}} 
\end{array}\right)
 \frac{v_{\scriptscriptstyle 0}}{\sqrt{2}}
-\,(\, \mu_{\scriptscriptstyle \za}^*
\lambda_{{\scriptscriptstyle \za\zb}k}\, ) \, 
\frac{v_{\scriptscriptstyle u}}{\sqrt{2}} \; .
\eeqa \normalsize
For the neutral scalars, we have explicitly
\beq \label{MSN}
{\cal M}_{\!\scriptscriptstyle S}^2 =
\left( \begin{array}{cc}
{\cal M}_{\!\scriptscriptstyle SS}^2 &
{\cal M}_{\!\scriptscriptstyle SP}^2 \\
({\cal M}_{\!\scriptscriptstyle SP}^{2})^{\!\scriptscriptstyle T} &
{\cal M}_{\!\scriptscriptstyle PP}^2
\end{array} \right) \; ,
\eeq
where the scalar, pseudo-scalar, and mixing parts are given by
\beqa
{\cal M}_{\!\scriptscriptstyle SS}^2 &=&
\mbox{Re}({\cal M}_{\!\scriptscriptstyle {\phi}{\phi}\dag}^2)
+ {\cal M}_{\!\scriptscriptstyle {\phi\phi}}^2 \; ,
\nonumber \\
{\cal M}_{\!\scriptscriptstyle PP}^2 &=&
\mbox{Re}({\cal M}_{\!\scriptscriptstyle {\phi}{\phi}\dag}^2)
- {\cal M}_{\!\scriptscriptstyle {\phi\phi}}^2 \; ,
\nonumber \\
{\cal M}_{\!\scriptscriptstyle SP}^2 &=& -
\mbox{Im}({\cal M}_{\!\scriptscriptstyle {\phi}{\phi}\dag}^2) \; ,
\label{lastsc}
\eeqa
respectively, with 
\small\beqa \label{Mpp}
{\cal M}_{\!\scriptscriptstyle {\phi\phi}}^2 
 &=& {1\over 2} \, M_{\!\scriptscriptstyle Z}^2\,
\left( \begin{array}{ccc}
 \sin\!^2\! \beta   &  - \cos\!\beta \, \sin\! \beta
 &  \quad 0_{\scriptscriptstyle 1 \times 3} \\
 - \cos\!\beta \, \sin\! \beta \!\! & \!\! \cos\!^2\! \beta 
 &  \quad 0_{\scriptscriptstyle 1 \times 3} \\
0_{\scriptscriptstyle 3 \times 1} &  0_{\scriptscriptstyle 3 \times 1} 
 & \quad 0_{\scriptscriptstyle 3 \times 3} 
\end{array} \right) ,
\eeqa \normalsize
{\normalsize and}
\beqa 
{\cal M}_{\!\scriptscriptstyle {\phi}{\phi}^\dag}^2 
 &=&  {\cal M}_{\!\scriptscriptstyle {\phi\phi}}^2 +
 \left(  \begin{array}{cc}
\tilde{m}_{\!\scriptscriptstyle H_{\!\scriptscriptstyle u}}^2
+ \mu_{\!\scriptscriptstyle \za}^* \mu_{\scriptscriptstyle \za}
  -\frac{1}{2} \, M_{\!\scriptscriptstyle Z}^2\, \cos\!2 \beta
& - (B_\za) \\
- (B_\za^*) &
\tilde{m}_{\!\scriptscriptstyle {L}}^2 
+ (\mu_{\!\scriptscriptstyle \za}^* \mu_{\scriptscriptstyle \zb})
+ \frac{1}{2} \, M_{\!\scriptscriptstyle Z}^2\, \cos\!2 \beta
\end{array}  \right) \; .
\label{Mp}
\eeqa \normalsize
Note that $\tilde{m}_{\!\scriptscriptstyle {L}}^2$ here is a $4\times 4$
matrix of soft masses for the $L_\za$, and $B_\za$'s are the corresponding bilinear
soft terms of the $\mu_{\scriptscriptstyle \za}$'s.
$A^{\!{\scriptscriptstyle E}}$ is just the $3\times 3$ R-parity conserving
leptonic $A$-term. There is no contribution from the admissible RPV $A$-terms
under the SVP. Also, we have used
$m_{\!\scriptscriptstyle L} \equiv \mbox{diag} \{\,0, m_{\!\scriptscriptstyle E}\,\}
\equiv \mbox{diag} \{\,0, m_{\scriptscriptstyle 1}, m_{\scriptscriptstyle 2}, 
m_{\scriptscriptstyle 3}\,\}$.

The RPV contributions to the charged, as well as neutral, scalar masses and 
mixings give rise to new terms in electron EDM and 
$\mu \to e\, \gamma$\cite{as7,as11,cch1,cch2}, as well as neutrino masses diagrams 
that had been largely overlooked\cite{as1,as5}. From our extensive analytical
and numerical study\cite{as7}, a brief summary of results are shown in Table~1.
\begin{table}[h] \begin{center}
\caption{\small \label{table2}
Illustrative bounds on  combinations 
of  R-parity violating parameters
from $\mu\to e\,\gamma$.} 
\begin{tabular}{|lr|}\hline 
\ \ 
$\frac{|{\mu_{\scriptscriptstyle 3}^*}\,{\lambda_{\scriptscriptstyle 321}}|}
{|\mu_{\scriptscriptstyle 0}|}\;, \;\;\;
\frac{|{\mu_{\scriptscriptstyle 1}^*}\,{\lambda_{\scriptscriptstyle 121}}|}
{|\mu_{\scriptscriptstyle 0}|}\;, \;\;\;
\frac{|{\mu_{\scriptscriptstyle 3}}\,{\lambda_{\scriptscriptstyle 312}^*}|}
{|\mu_{\scriptscriptstyle 0}|}\;, \;\;\; 
\mbox{or} \;\;\;
\frac{|{\mu_{\scriptscriptstyle 2}}\,{\lambda_{\scriptscriptstyle 212}^*}|}
{|\mu_{\scriptscriptstyle 0}|}\; \;\;\;$ & 
$< 1.5 \times 10^{-7}$ \ \ 
\\ \ \
$\frac{|\mu_{\scriptscriptstyle 1}^*\, \mu_{\scriptscriptstyle 2}|}
{|\mu_{\scriptscriptstyle 0}|^2}$	&	$ < 0.53 \times 10 ^{-4}$ \ \
\\ \ \
$|\lambda_{\scriptscriptstyle 321} \lambda^*_{\scriptscriptstyle 131}|\;, \;\;\;
|\lambda_{\scriptscriptstyle 322} \lambda^*_{\scriptscriptstyle 132}|\;, \;\;\; 
\mbox{or} \;\;\;
|\lambda_{\scriptscriptstyle 323} \lambda^*_{\scriptscriptstyle 133}|$      
& $<2.2 \times 10^{-4}$ \ \ 
\\ 
\ \
$|\lambda^*_{\scriptscriptstyle 132} \lambda_{\scriptscriptstyle 131}|\;, \;\;\;
|\lambda^*_{\scriptscriptstyle 122} \lambda_{\scriptscriptstyle 121}|\;, \;\;\; 
\mbox{or} \;\;\;
|\lambda^*_{\scriptscriptstyle 232} \lambda_{\scriptscriptstyle 231}|$      
& $<1.1 \times 10^{-4}$ \ \ 
\\ 
\ \  
$\frac{|B_{3}^*\,\lambda_{\scriptscriptstyle 321}|}{|\mu_{\scriptscriptstyle 0}|^2}
\;, \;\;\;
\frac{|B_{1}^*\,\lambda_{\scriptscriptstyle 121}|}{|\mu_{\scriptscriptstyle 0}|^2}
\;, \;\;\;
\frac{|B_{3}\,\lambda_{\scriptscriptstyle 312}^*|}{|\mu_{\scriptscriptstyle 0}|^2}
\;, \;\;\; \mbox{or} \;\;\;
\frac{|B_{2}\,\lambda_{\scriptscriptstyle 211}^*|}{|\mu_{\scriptscriptstyle 0}|^2}$
&  $<2.0\times 10^{-3}$ \ \ 
\\ 
\ \ 
$\frac{|B_1^* \, \mu_{\scriptscriptstyle 2}|}{|\mu_{\scriptscriptstyle 0}|^3}$
 & $< 1.1\times 10^{-5}$  \ \ 
\\ \hline
\end{tabular}\end{center}
\end{table}

\section*{Acknowledgments}
The author thanks M. Bisset, K. Cheung, S.K. Kang,
Y.-Y. Keum, C. Macesanu, and L.H. Orr, for collaborations on the subject, 
and colleagues at Academia Sinica for support.


\begin{thebibliography}{99}

\bibitem{ru}
M. Bisset, O.C.W. Kong, C. Macesanu, and L.H. Orr,
Phys. Lett. {\bf B430}, 274 (1998); 
Phys. Rev. {\bf D62},  {\it 035001} (2000).
\bibitem{as8}
O.C.W. Kong, {IPAS-HEP-k008},
{\it manuscript in preparation}.
\bibitem{as5}
O.C.W. Kong, JHEP {\bf 0009}, {\it 037} (2000).
\bibitem{ok}
O.C.W. Kong, Mod. Phys. Lett. {\bf A14},  903  (1999).
\bibitem{as1}
K. Cheung and O.C.W. Kong,  
Phys. Rev. {\bf D61},  {\it 113012} (2000).
\bibitem{as9}
S.K. Kang and O.C.W. Kong, {IPAS-HEP-k009},
{\it manuscript in preparation}.
\bibitem{AL}
See also
A. Abada and M. Losada,  hep-ph/9908352;
S. Davidson and M. Losada, JHEP {\bf 0005}, {\it 021} (2000);
hep-ph/0010325.
\bibitem{as4}
Y.-Y. Keum and O.C.W. Kong,  Phys. Rev. Lett. {\bf 86}, 393 (2001).
\bibitem{as6}
Y.-Y. Keum and O.C.W. Kong, Phys. Rev. {\bf D63}, {\it 113012} (2001).
\bibitem{kk}
O.C.W. Kong {\it et.al.}, {\it work in progress}.
\bibitem{cch1}
See also 
K. Choi, E.J. Chun, and K. Hwang,  Phys. Rev. {\bf D63},  {\it 013002} (2001).
\bibitem{as7}
K. Cheung and O.C.W. Kong, 
hep-ph/0101347, {\it  submitted to Phys. Rev. D}.
\bibitem{as11}
K. Cheung, Y.-Y. Keum, and O.C.W. Kong, {IPAS-HEP-k011},
{\it manuscript in preparation}.
\bibitem{cch2}
See also
K. Choi, E.J. Chun, and K. Hwang, Phys. Lett. {\bf B488}, 145 (2000).
\end{thebibliography}
\end{document}